\documentclass[a4paper,11pt]{article}
\usepackage{pos}

\title{Four-loop scattering amplitudes journey into the forest}

\author*[a,b,c]{Selomit Ram\'irez-Uribe}
\author[b]{Roger J. Hern\'andez-Pinto}
\author[a]{Andr\'es E. Renter\'{\i}a-Olivo}
\author[a]{Germ\'an Rodrigo}
\author[a,d]{German F. R. Sborlini}
\author[e]{William~J.~Torres~Bobadilla}
\author[a]{Luiz Vale Silva}

\affiliation[a]{Instituto de F\'isica Corpuscular, Universitat de Val\`encia -- Consejo Superior de Investigaciones Cient\'ificas, 
Parc Cient\'ific, E-46980 Paterna, Valencia, Spain.}

\affiliation[b]{Facultad de Ciencias F\'isico-Matem\'aticas,
Universidad Aut\'onoma de Sinaloa, 
Ciudad Universitaria, CP 80000 Culiac\'an, Mexico.}

\affiliation[c]{Facultad de Ciencias de la Tierra y el Espacio,
Universidad Aut\'onoma de Sinaloa, Ciudad Universitaria, CP 80000 Culiac\'an, Mexico.}

\affiliation[d]{Deutsches Elektronen-Synchrotron DESY, Platanenallee 6, 15738 Zeuthen, Germany.}

\affiliation[e]{Max-Planck-Institut für Physik, Werner-Heisenberg-Institut, 80805 München, Germany.}

\emailAdd{norma.selomit.ramirez@ific.uv.es}
\emailAdd{roger@uas.edu.mx}
\emailAdd{andres.renteria@ific.uv.es}
\emailAdd{german.rodrigo@csic.es}
\emailAdd{german.sborlini@desy.de}
\emailAdd{torres@mpp.mpg.de}
\emailAdd{luizva@ific.uv.es}

\abstract{We present an overview of the analysis of the multiloop topologies that appear for the first time at four loops and the assembly of them in a general expression, the N$^4$MLT universal topology. Based on the fact that the Loop-Tree Duality enables to open any scattering amplitude in terms of convolutions of known subtopologies, we go through the dual representation of the universal N$^4$MLT topology and the manifestly causal representation. Additionally, we expose the application of a quantum algorithm as an alternative methodology to identify the causal singular configurations of multiloop Feynman diagrams.}

\FullConference{%
  *** The European Physical Society Conference on High Energy Physics (EPS-HEP2021), ***\\
  *** 26-30 July 2021 ***\\
  *** Online conference, jointly organized by Universität Hamburg and the research center DESY ***
}

\def\ii{\imath 0}
\def\Eq#1{Eq.~(\ref{#1})}
\def\beq{\begin{equation}}
\def\eeq{\end{equation}}
\def\bea{\begin{eqnarray}}
\def\eea{\end{eqnarray}}
\def\nn{\nonumber}
\def\qon#1{q_{#1,0}^{(+)}}

\def\ket#1{|{#1}\rangle}
\def\bra#1{\langle{#1}|}
\def\id{\boldsymbol I}

\begin{document}
\maketitle
\section{Introduction}
A critical challenge in perturbative Quantum Field Theory is the description of quantum fluctuations at high-energy scattering processes by the calculation of multiloop scattering amplitudes. Aiming for improving the efficiency of these computations, we work in a framework based on the Loop-Tree Duality (LTD) ~\cite{Catani:2008xa,Bierenbaum:2010cy,Bierenbaum:2012th,Tomboulis:2017rvd,Runkel:2019yrs,Capatti:2019ypt,Verdugo:2020kzh}.

The LTD is a novel method that opens any loop diagram to a forest of non-disjoint trees. One of the most important features of LTD is the distinction between physical and unphysical singularities at integrand level~\cite{Buchta:2014dfa,Aguilera-Verdugo:2019kbz}. Besides this, LTD has other interesting characteristics: for instance, in numerical implementations the number of integration variables is independent of the number of external legs~\cite{Buchta:2015wna,Buchta:2015xda,Driencourt-Mangin:2019yhu,Capatti:2019edf,Jurado:2017xut}, it straightforward provides asymptotic expansions~\cite{Beneke:1997zp,Driencourt-Mangin:2017gop,Plenter:2019jyj,Plenter:2020lop}, and promising local renormalization approaches~\cite{Driencourt-Mangin:2019aix,Prisco:2020kyb}.
Furthermore, an important associated development was the proposal of computing cross sections directly in four space-time dimensions through the so-called, Four Dimensional Unsubtraction (FDU)~\cite{Hernandez-Pinto:2015ysa,Sborlini:2016gbr,Sborlini:2016hat,Driencourt-Mangin:2019sfl}.

A crucial achievement was the conjecture that LTD leads to very compact and manifestly causal representations of scattering amplitudes to all orders~\cite{Verdugo:2020kzh, snowmass2020}.
This statement was proven for a series of multiloop topologies, the maximal (MLT), next-to-maximal (NMLT) and next-to-next-to-maximal (N$^2$MLT) loop topologies.
Their analytic dual representations are implicitly free of unphysical singularities, and their causal structure is interpreted in terms of entangled causal thresholds~\cite{Aguilera-Verdugo:2020kzc,Sborlini:2021owe,Bobadilla:2021pvr}. We present here the analysis of the topologies that appear for the fist time at four loops~\cite{Ramirez-Uribe:2020hes}.

In the causal representation context, a problem to be solved through the LTD framework is the identification of all internal causal configurations among the $N=2^n$ potential solutions, where $n$ is the number of internal Feynman propagators. This is a problem that can be handled with a quantum computing approach, applying Grover's algorithm~\cite{Grover:1997fa} for querying over unstructured databases~\cite{Boyer:1996zf}. The first benchmark application of a quantum algorithm to Feynman loop integrals has been presented in Ref.~\cite{Ramirez-Uribe:2021ubp}, where the causal singular configurations of multiloop Feynman diagrams were unfolded.

\section{Loop-tree duality}
An arbitrary $L$-loop scattering amplitude with $P$ external legs, $\{p_j\}_P$, is written in the Feynman representation as an integral in the Minkowski space of the $L$ loop momenta, $\{\ell_s\}_L$,
\begin{align}\label{eq:LamplitudeN}
\mathcal{A}_F^{(L)}(1,\ldots, n) = \int_{\ell_1, \ldots, \ell_L}  \mathcal{N}( \{ \ell_s\}_L,  \{ p_j\}_P) \, G_F(1,\ldots, n)\, ,
\end{align}
where the integration measure in dimensional regularization~\cite{Bollini:1972ui, tHooft:1972tcz} reads $
\int_{\ell_s} = -\imath \mu^{4-d} \int \frac{d^d\ell_s}{(2\pi)^d}$, 
with $d$ the number of space-time dimensions.
The integrand in the Feynman representation in \Eq{eq:LamplitudeN}, is composed by a numerator depending on the particles and the interactions involved, and  
$G_F(1,\ldots, n) = \prod_{i\in 1\cup\ldots \cup n} \left( G_F(q_i) \right)^{a_i}$ 
denoting the product of Feynman propagators, which are grouped by sets according to their dependence on a specific loop momentum or linear combination of loop momenta. The Feynman propagators written in terms of the positive on-shell energy component are $G_F(q_i) = \left(q_{i,0}^2-\left(q_{i,0}^{(+)}\right)^2 \right)^{-1}$, with $q_{i,0}^{(+)}=\sqrt{{\bf q}_i^2+m_i^2-\ii}$ . 

To obtain the LTD representation, we integrate out one degree of freedom per loop applying the Cauchy's residue theorem. 
Considering the case of multiloop scattering amplitudes, 
the LTD representation is written in terms of nested residues~\cite{Verdugo:2020kzh,Aguilera-Verdugo:2020fsn}
\begin{align}
\label{eq:nested}
&\mathcal{A}_D^{(L)}(1,\ldots, r; r+1,\dots, n)  
=-2\pi \imath \sum_{i_r \in r} {\rm Res} (\mathcal{A}_D^{(L)}(1, \ldots, r-1;r, \ldots, n), {\rm Im}(\eta\cdot q_{i_r})<0)\, , 
\end{align}
starting from $r=1$, which corresponds to the Feynman representation, \Eq{eq:LamplitudeN}.
All sets before the semicolon contain one on-shell propagator, the remaining sets located after the semicolon have all the propagators off shell. To evaluate the residue, we select the poles with negative imaginary components through the implementation of the future-like vector $\eta$ indicating the component of the loop momenta to be integrated. The most convenient choice is $\eta^{\mu}=(1,{\bf 0})$, allowing to go in the integration domain from a Minkowsky space to an Euclidean space.  

\section{N$^4$MLT universal topology}
The topologies that first appear at four loops are described by multiloop diagrams with $L+4$ and $L+5$ sets of propagators which correspond to the next-to-next-to-next-to maximal loop topology (N$^3$MLT) and next-to-next-to-next-to-next-to maximal loop topology (N$^4$MLT). The N$^4$MLT family consists of three main topologies and includes in a natural way all N$^{k-1}$MLT configurations with $k\le 4$.
A unified description of these topologies is achieved by interpreting them as the $t$-, $s$- and $u$-kinematic channels, of the N$^4$MLT {\it universal topology}.

\begin{figure}[t!]
\includegraphics[scale=0.6]{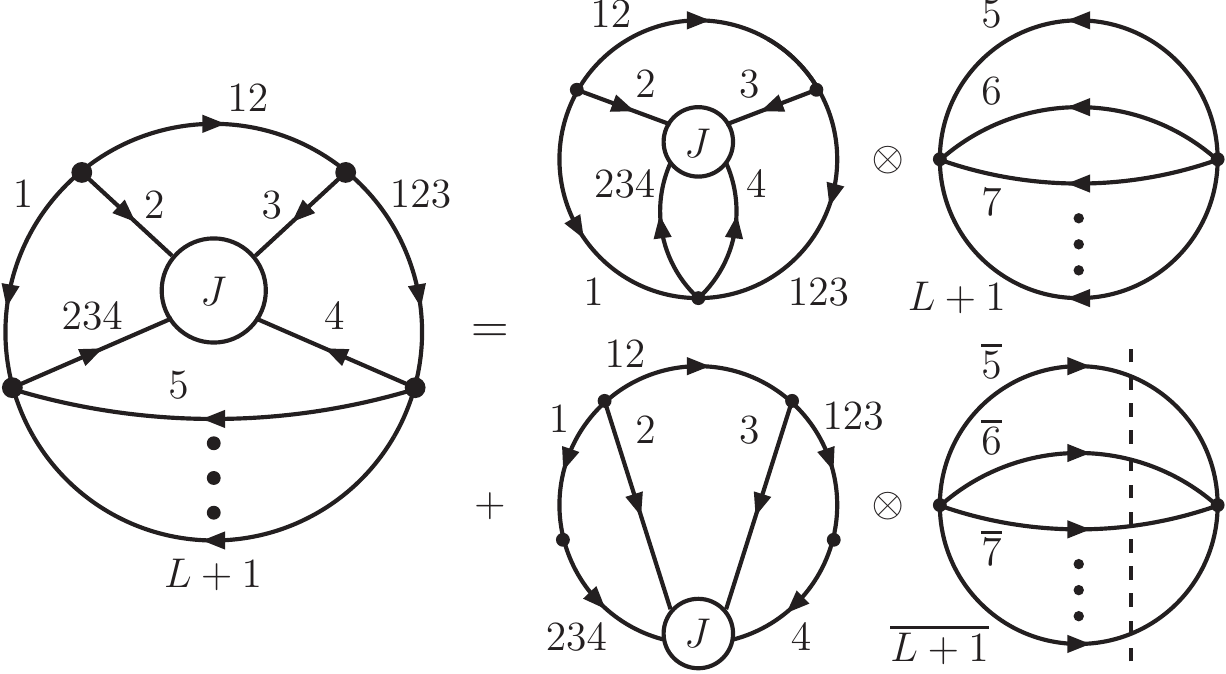}
\centering
\caption{Diagrammatic representation for the factorized opening of the multiloop 
N$^4$MLT {\it universal topology}. Only the on-shell cut of the last MLT-like subtopology 
with reversed momentum flow is shown.
\label{fig:master}}
\end{figure}
The three topologies differ only in one set of propagators. The sets $23$, $34$, and $24$ are considered as the distinctive key to each of the channels where the momenta of their propagators are taken as different linear combinations of $\ell_2$, $\ell_3$ and $\ell_4$. The application of the nested residues to the universal topology gives the dual opening as, 
\begin{align}
\label{eq:master}
{\cal A}^{(L)}_{\rm N^4MLT} & (1, \ldots, L+1, 12, 123, 234, J) \nn \\
& =  {\cal A}^{(4)}_{\rm N^4MLT} (1, 2, 3, 4, 12, 123, 234, J) \otimes {\cal A}_{\rm MLT}^{(L-4)} (5, \dots, L+1) \nn \\
& + {\cal A}^{(3)}_{\rm N^2MLT} (1\cup 234, 2, 3, 4\cup 123, 12, J) \otimes {\cal A}_{\rm MLT}^{(L-3)} (\overline 5, \dots, \overline{L+1} )~, 
\end{align}
expressed in a factorized form depicted as in Fig.~\ref{fig:master} and valid for any internal configuration. We call it universal opening identity given the fact that it is the only master expression needed to open to nondisjoint trees any scattering amplitude of up to four loops. The current $J$ is defined as $J=23\cup 34\cup 24$.

The terms ${\cal A}_{\rm MLT}^{(L-4)} (5, \dots, L+1)$ and ${\cal A}_{\rm MLT}^{(L-3)} (\overline 5, \dots, \overline {L+1})$ on the r.h.s. of \Eq{eq:master} are computed according to  Ref.~\cite{Verdugo:2020kzh,Aguilera-Verdugo:2020fsn}; 
the terms $\mathcal{A}^{(4)}_{\rm N^4MLT}$ and ${\cal A}^{(3)}_{\rm N^2MLT}$ consider all possible configurations with four and three on-shell conditions respectively, and are opened through a factorization identity which is written in terms of known subtopologies. These terms contemplate dual trees where all the propagators in $J$ remain off shell and for the contributions that characterize the $s$, $t$ or $u$ channel, the explicit expressions are presented in Ref.~\cite{Ramirez-Uribe:2020hes}.

\section{Causal LTD representations}
\label{sec:causal}

The confirmation of the causal conjecture for the N$^4$MLT family, follows the strategy proposed in \cite{Aguilera-Verdugo:2020kzc} and is applied to the multiloop N$^3$MLT, $t$, $s$ and $u$ channels. The configuration used for each topology is: one internal propagator in each loop set, four external momenta for N$^3$MLT and six external particles for $t$, $s$ and $u$ channels. Scalar integrals were considered given that they fully encode all the compatible causal matchings.

 A manifestly causal expression is found after the straightforward application of the universal opening in \Eq{eq:master} and adding all the dual terms together. Nevertheless, the numerator is a lengthy polynomial in the on-shell and external energies, therefore, to derive a more appropriate expression we reinterpret it in terms of entangled thresholds as defined in Ref.~\cite{Aguilera-Verdugo:2020kzc}.
 
A causal representation of the multiloop N$^3$MLT was analytically reconstructed in Ref.~\cite{Sborlini:2021owe,TorresBobadilla:2021ivx,Bobadilla:2021pvr} by matching all the combinations of four thresholds that are causally compatible to each other. There are thirteen causal denominators which represent potential singular configurations and they are constructed from sums of on-shell energies exclusively, i.e. they have the form $\lambda_p^\pm = \sum_{i\in p} q_{i,0}^{(+)} \pm k_{p,0}$. 

In the case of the N$^4$MLT family, all the entangled configurations involving five causal thresholds are considered. The causal representation of the $t$-channel depends on the causal denominators already defined for the N$^3$MLT configuration and nine extra causal denominators that depend on $\qon{23}$. For the $s$-channel, a clockwise rotation is applied to the $t$-channel; for the $u$-channel, besides a convenient substitution to the $t$-channel, three additional thresholds arise given the nonplanar context. All the details and specific results of the three N$^4$MLT causal representation are given in Ref.~\cite{Ramirez-Uribe:2020hes}.  

The main difference between the direct and causal LTD representations is the absence of noncausal singularities. The straightforward application of the nested residue generates multiple threshold singularities, nevertheless, with a clever analytical rearrangement, the absence of noncausal singularities is achieved and leads to a causal representation which is more stable numerically in all the integration domain \cite{Aguilera-Verdugo:2020kzc}.

\section{Causal quantum algorithm}

The idea to explore the application of quantum algorithms to Feynman loop integrals arise due to the implicit connection between a Feynman propagator and a qubit. Feynman propagators have two possible on-shell states which can be encoded in one qubit.  

Given the nature of the problem of identifying the causal configurations of selected topologies, a modified Grover's quantum algorithm is applied~\cite{Ramirez-Uribe:2021ubp}. The standard Grover's querying algorithm over unstructured databases starts from a uniform superposition, with a total of $N=2^n$ states which can be understood as the superposition of a winning state $\ket{w}$ and the orthogonal state $\ket{q_\perp}$, i.e.
\beq
\ket{q}= \frac{1}{\sqrt{N}} \sum_{x=0}^{N-1} \ket{x}~, \qquad  
\ket{q} =  \cos \theta \, \ket{q_\perp} + \sin\theta \, \ket{w}~,
\eeq
where $\ket{w}$ gathers all the causal solutions in an uniform superposition and $\ket{q_\perp}$ the noncausal states. The mixing angle is given by $\theta =  \arcsin \sqrt{r/N}$, with $r$ the number of causal solutions. 
The algorithm requires two operators, the oracle  and diffusion operators: $U_w = \id - 2\ket{w} \bra{w}$ and $U_q = 2 \ket{q} \bra{q} - \id$, respectively.
The oracle operator flips the state $\ket{x}$ if $x\in w$, $U_w \ket{x} = - \ket{x}$, and leaves it unchanged otherwise.
In the case of the diffusion operator, $U_q$ performs a reflection around the initial state $\ket{q}$ with the purpose of amplify the winning state probability. The iterative application of both operators $t$ times leads to 
\beq\label{eq:iteration}
(U_q U_w )^t \ket{q} = \cos \theta_t \, \ket{q_\perp } +  \sin \theta_t \, \ket{w}, \quad \theta_t = (2t +1) \, \theta. 
\eeq

Based on \Eq{eq:iteration}, the target is to get a final state such that the probability associated to the orthogonal state is considerable much smaller than the probability of the causal solutions. To achieve this, it is important to consider the range of $\theta$ because Grover's algorithm works well if $\theta \leq \pi/6$ $(r \leq N/4)$, but does not provide the desired amplitude amplification for greater angles. 

Studying the N$^4$MLT family we found that the causal configurations are nearly half of the total sates. To overcome this condition, we take advantage of the fact that given a causal solution, reversing all momentum flows is also a causal solution. 
Therefore, we can reduce the number of causal solutions to query by half, fixing one of the propagators set flow and assuming that only one of its states contributes to the winning set.

The quantum algorithm proposed needs three registers for implementation, together with an extra qubit used as marker by the Grover's oracle.  
The first register, $q_i$, encodes the state of the propagators.
The qubit $q_i$ is in the state $\ket{1}$ if the momentum flow of the corresponding line is oriented as the initial assignment and in $\ket{0}$ if it is in the opposite direction. 
The second register stores the Boolean clauses, $c_{ij}$,
probing whether or not two adjacent qubits are oriented in the same direction. 
These binary clauses are defined as
$c_{ij} \equiv (q_i = q_j)$, $\bar c_{ij} \equiv (q_i \ne q_j)$ with $i,j \in\{0, \ldots, n-1\}~$.
The third register, $a_k(\{c_{ij}\},\{\bar c_{ij}\})$, encodes the loop clauses that probe if all the qubits in each subloop form a cyclic circuit. 

\begin{figure}[t]
\includegraphics[width=170px]{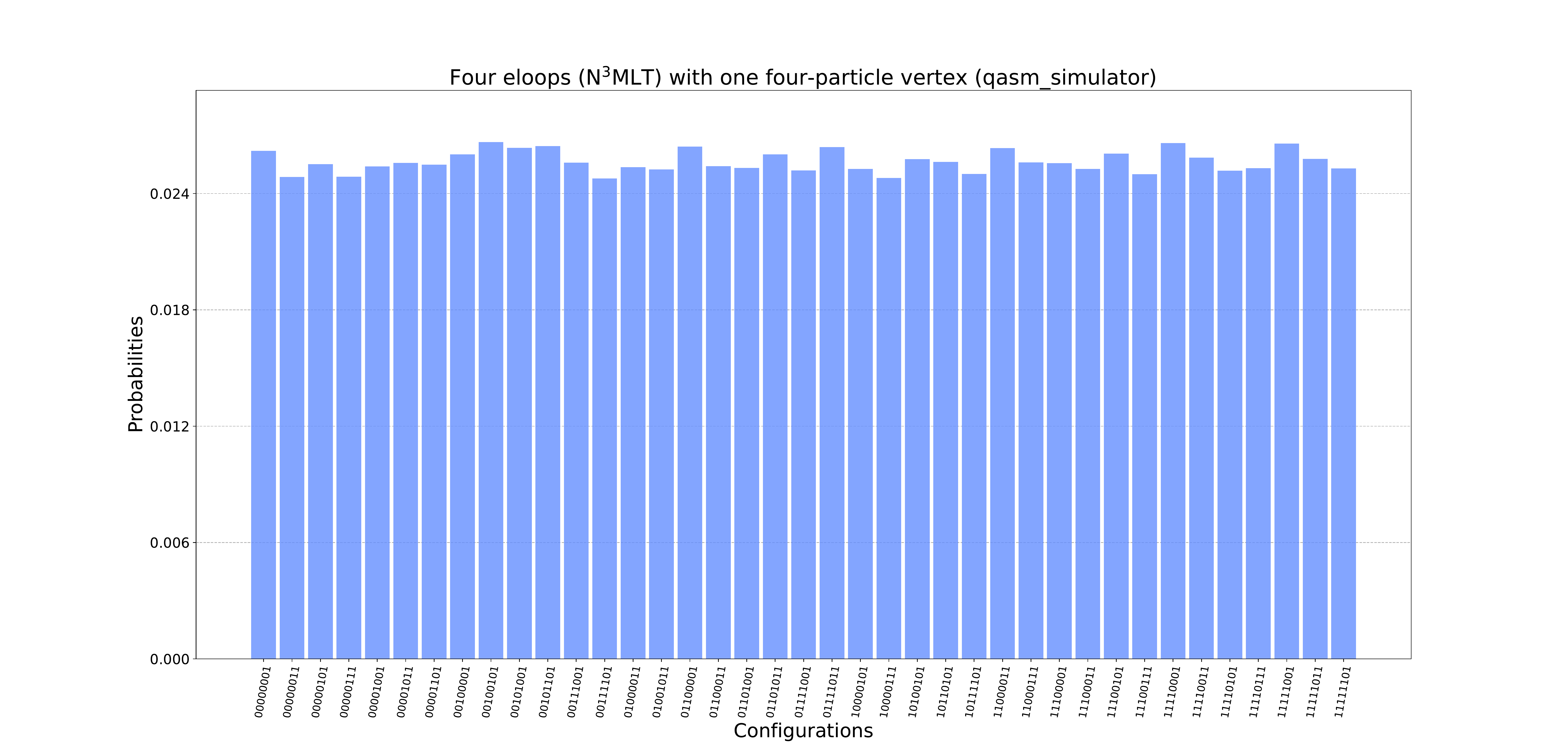}
\includegraphics[width=260px]{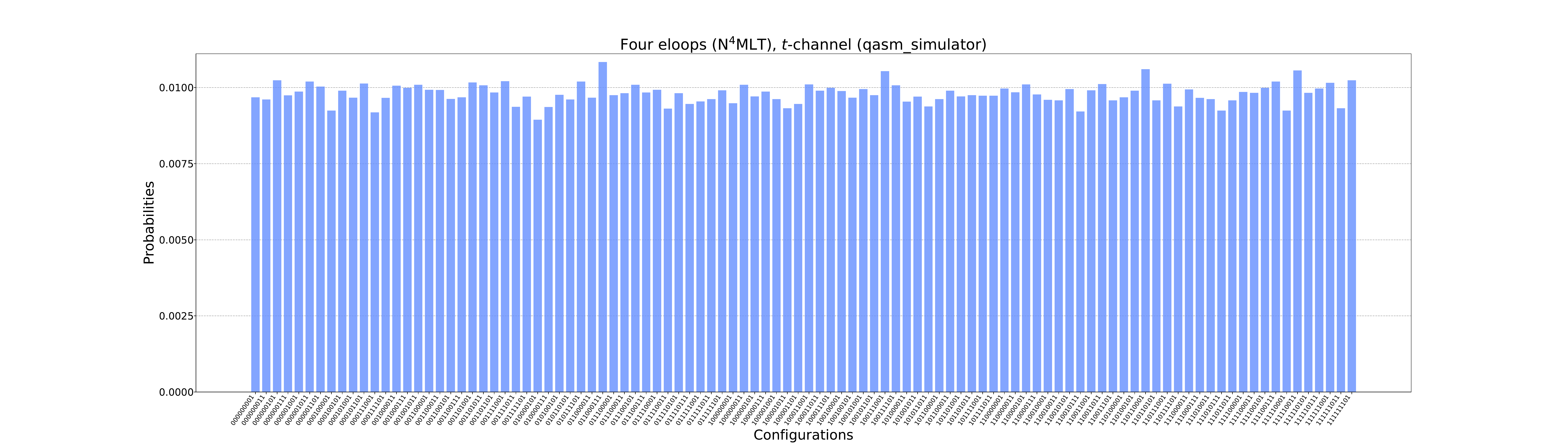}
\caption{Probabilities of causal states for four-loop configurations.
On the left, N$^3$MLT; on the right the $t$-channels of N$^4$MLT with $n_i=1$. 
The number of selected states is $39/256$ and $102/512$, respectively.
\label{fig:prob4eloops}}
\end{figure}

The causal quantum algorithm is implemented as follows. 
The initial uniform superposition is obtained by applying Hadamard gates to each of the qubits in the $q$-register, $\ket{q} = H^{\otimes n} \ket{0}$. The $\ket{c}$ and $\ket{a}$ registers are initialized to $\ket{0}$ while the qubit which is used as Grover's marker is initialized to the Bell state $\ket{out_0} = \ket{-} = \left(\ket{0} - \ket{1}\right)/\sqrt{2}$. 
Each binary clause $\bar c_{ij}$ requires two CNOT gates operating between two qubits in the $\ket{q}$
register and one qubit in the $\ket{c}$ register. An extra XNOT gate acting on the corresponding qubit in 
$\ket{c}$ is needed to implement a $c_{ij}$ binary clause. 

The oracle is defined as $
U_w \ket{q} \ket{c} \ket{a} \ket{out_0} = \ket{q} \ket{c} \ket{a} \ket{out_0 \otimes f(a,q)}$ ,
with $\ket{out_0 \otimes 0} = \ket{out_0}$ and $\ket{out_0 \otimes 1} = - \ket{out_0}$. Therefore, if all the causal conditions are satisfied, $f(a,q) = 1$ and the corresponding states are marked;
otherwise $f(a,q) = 0$. After the marking, the oracle operations are applied in inverse order. Then, the diffusion $U_q$ is applied to the register $\ket{q}$.
The diffuser used is the one described in the IBM Qiskit website (\texttt{https://qiskit.org/}). Also, we used the IBM's quantum simulator provided by the open source Qiskit framework, having as a current upper limit of $32$ qubits.

This algorithm requires one single iteration for all the cases analyzed. To illustrate the performance of the quantum algorithm, we show in Fig.~\ref{fig:prob4eloops}
the probability of the winning states obtained in the IBM's Qiskit simulator related to the N$^3$MLT and the $t$-channel of the N$^4$MLT. Despite the complexity of these four-loop diagrams, all the causal configurations were successfully identified, as the algorithm noticeably enhanced their probabilities with respect to the noncausal configurations~\cite{Ramirez-Uribe:2021ubp}.

\section{Conclusions}

We have presented a unified description and representation of scattering amplitudes up to four loops through the universal N$^4$MLT topology. The LTD was applied to this topological family to all orders, managing to get a LTD representation which exhibits a recursive form in terms of simpler topologies. 

An additional step has been taken in the LTD framework, the causal LTD representation is explicitly found and reinterpreted in terms of entangled causal thresholds. This representation allows to work with favorable conditions, the absence of noncausal thresholds, enabling to get more efficient numerical evaluation of multiloop scattering amplitudes.

Based on quantum computing, we proposed a novel strategy to identify causal configurations of  Feynman integrals. The problem has been restated for the proper application of Grover's quantum algorithm and the causal singular configurations have been efficiently unfolded for the N$^3$MLT and N$^4$MLT topologies.

\section*{Acknowledgements}
This work is supported by the Spanish Government (Agencia Estatal de Investigaci\'on)  and ERDF funds from European Commission (Grant No. PID2020-114473GB-I00),  Generalitat  Valenciana (Grant No. PROMETEO/2021/071) and the  COST  Action  CA16201  PARTICLEFACE.  
SRU acknowledges support from CONACyT and  Universidad  Aut\'onoma  de  Sinaloa;
RJHP from the Project No. A1-S-33202 (Ciencia B\'asica), Ciencia de Frontera 2021-2042 and Sistema Nacional de Investigadores; 
AERO from the Spanish Government (PRE2018-085925).

\end{document}